\documentclass[a4paper]{article}

\usepackage{icrc2013}
\usepackage{amssymb}

\title{Observations of TeV binary systems with the H.E.S.S. telescope}

\shorttitle{H.E.S.S. binaries}

\authors{

Pol Bordas$^{1}$,  H\'{e}l\`{e}ne Laffon$^{2}$, Mathieu de Naurois$^{3}$, Stefan Ohm$^{4}$, Emma de O\~{n}a Wilhelmi$^{5}$,  Iurii Sushch$^{6, 7, 8}$,  Francesca Volpe$^{5}$, and  V\'{i}ctor Zabalza$^{5}$, for the H.E.S.S.
Collaboration

}

\afiliations{
$^1$Institut f{\"u}r Astronomie und Astrophysik, Universit{\"a}t T{\"u}bingen, Sand 1, D-72076 T{\"u}bingen, Germany\\
$^2$Universit\'{e} Bordeaux 1, CNRS/IN2P3, Centre d’\'{E}tudes Nucl\'{e}aires de Bordeaux Gradignan, 33175 Gradignan, France \\
$^3$Laboratoire Leprince-Ringuet, Ecole Polytechnique, CNRS/IN2P3, F-91128 Palaiseau, France \\
$^4$Department of Physics and Astronomy, The University of Leicester, University Road, Leicester, LE1 7RH, United Kingdom \\
$^5$Max-Planck-Institut f\"{u}r Kernphysik, P.O. Box 103980, D 69029 Heidelberg, Germany \\
$^6$Centre for Space Research, North-West University, Potchefstroom Campus, Potchefstroom, South Africa \\
$^7$Astronomical Observatory of Ivan Franko National University of Lviv, Lviv, Ukraine\\
$^8$Humboldt Universit\"{a}t zu Berlin, Institut f\"{u}r Physik, Berlin, Germany\\
}

\email{pol.bordas@uni-tuebingen.de}

\abstract{Recent observations of binary systems obtained with the High Energy Stereoscopic System of Cherenkov telescopes are providing
crucial information on the physics of relativistic outflows and the engines powering them. We report here on new  H.E.S.S. results on
HESS~J0632+057, PSR~B1259-63/LS~2883, Eta Carinae and the recently discovered source HESS J1018-589. Despite the high-quality data
obtained in the last years through both ground and space-based gamma-ray detectors, many questions on the mechanisms that permit binary
systems to emit at gamma-rays remain open. In particular, it is becoming apparent that emission at high and very-high energies is
uncorrelated in some gamma-ray binary systems, with bright GeV flares not observed at TeV energies (e.g. PSR B1259-63), and sources
periodically detected at VHEs which are lacking its HE counterpart (e.g. HESS~J0632+057). Our results mainly confirm the predictions
derived previously for the studied sources, but unexpected results are also found in a few cases, which are discussed in the context of
contemporaneous observations at lower energies.}


\begin{document}
\maketitle

%Begin a section.
\section{Introduction}
%-------------------------------------------------------------------------------------------------------------------

Gamma-ray binary systems are  X-ray binaries displaying the peak of their non-thermal energy flux at gamma-ray energies. The total number of known systems is still low, owing to the required sensitivity of instruments operating in the High- and Very High Energy (HE, 100~MeV~$\leq E \leq$~100 GeV; VHE, $E > $100~GeV) gamma-ray domains, which became only available in recent years with the third generation of Cherenkov telescopes as well as with improved pair-conversion space-based detectors. Gamma-ray binaries are further characterized by displaying variable and in most cases orbitally modulated emission.  All systems known so far are high-mass X-ray binaries, consisting of a compact object orbiting around a massive star of O or Be type. Implicitly, the dense photon and/or matter densities provided by the powerful companions are required to generate gamma-rays, through Inverse Compton (IC) and/or hadronic interactions. In addition, young and bright stars are mainly lying on the Galactic plane, where therefore all known gamma-ray binaries are also found. Regarding the compact object, the presence of a neutron star has been clearly identified only in the gamma-ray binary PSR~B1259-63/LS~2883 \cite{Aharonian2005a}, whereas for the systems LS~5039~\cite{Aharonian2005b}, LS~I~+61$^{\circ}$~303 \cite{Albert2009, Acciari2011} and the recently discovered sources HESS~J0632+057 \cite{Aharonian2007, Hinton2009} and 1FGL~J1018.6-5856 \cite{Fermi2012, HESS2012b}, the nature of the compact object is still unknown. In the case of Cyg~X-3 \cite{Abdo2009b, Tavani2009a, Aleksic2010} and Cyg X-1 \cite{Albert2007, Sabatini2010}, a black hole scenario seems to be preferred. Interestingly, neither a neutron star or a black hole are present in the HE gamma-ray emitting binary system Eta~Carinae \cite{Abdo2009a, Farnier2011}.

\vspace{0.5cm}

\noindent We summarize in these proceedings the recent H.E.S.S. observations of the systems PSR~B1259-63/LS~2883,  HESS~J0632+057, HESS~J1018-589 (with its possible association to 1FGL~J1018.6-5856) and Eta~Carinae. The reader is referred to a more complete description of the observations and the data-analysis as well as on the discussion of the obtained results shown in \cite{HESS2013, Bordas-Meier-2012, HESS2012a} and \cite{HESS2012b} for each of those sources, respectively. 

%\noindent The physical mechanisms leading to the observed $\gamma$-ray emission these systems are under debate. Two major scenarios
%have been put forward so far, based on accretion- or relativistic winds powered engines. In the first case, efficient particle
%acceleration is known to take place in X-ray binary systems displaying relativistic jets (\cite{Mirabel-1994}), which
%could extend to the gamma-ray domain. In the second case, particle acceleration in shocks created by the collision of a rotation
%powered pulsar wind (or a non-degenerated star as in Eta~Carinae) with the equatorial disk or wind of the companion star could also
%give place to high-energy radiation \cite{Maraschi-1981, Dubus-2006}. In any case, the interaction between relativistic outlfolws, and
%the presence of strong magnetic and radiation fields form a complex  environment for acceleration, radiation and  absorption processes
%(see e.g. \cite{Bednarek-2011a}), which may be ultimately responsible for the variety of emission patterns observed in these systems. 

\section{Observations}
\subsection{PSR~B1259-63/LS~2883}
%-------------------------------------------------------------------------------------------------------------------

PSR~B1259-63/LS~2883 was originally discovered in a high-frequency radio survey devoted to the detection of young and short-period pulsars \cite{Johnston-1992}. It consists of a rapidly rotating neutron star with a spin period of about 48 ms and a spin-down luminosity $\sim 10^{35}$~erg~s$^{-1}$, which orbits a massive Be star every $\sim 3.4$~years in a highly eccentric ($e = 0.87$) orbit. The Be companion displays a  dense equatorial disk, which seems to be inclined with respect to the pulsar orbital plane \cite{Negueruela-2011}. The pulsar is therefore forced to cross twice the Be disk in each orbit and, given the high eccentricity, both stars are found at relatively close distances during periastron passage. PSR~B1259-63/LS~2883 is a rather well-observed system and since it discovery it has been studied at all energy bands. In the TeV domain, the system was observed by H.E.S.S. during the 2004 and 2007 periastron passages, at slightly different orbital phases, providing a highly significant detection at VHEs on both occasions \cite{Aharonian2005b,Aharonian-2009}.  

New H.E.S.S. observations were scheduled to cover the 2010/2011 periastron passage. However, in this occasion the source was not visible for H.E.S.S. before and at the periastron passage. In addition, observations were partially hampered by bad weather conditions, resulting in a total data-set of $\sim6$~h of livetime after the quality selection cuts \cite{HESS2013}. These new observations partially overlapped with the start time of a strong gamma-ray flare reported by the {\it{Fermi}}-LAT, about 30~d after periastron. This HE flare lasted for about 7 weeks, displaying an average flux $\sim$10 times higher than that close to the periastron, and implied a very high efficiency for the conversion of the pulsar rotational energy into gamma-rays, approaching a 100\% level \cite{Abdo-2011}.

The new H.E.S.S. observations resulted in a clear detection of PSR~B1259-63/LS~2883, at a $\sim 11.5 \sigma$ significance level, and both the flux level and the spectral properties are consistent with those obtained in previous periastron observations \cite{HESS2013}. No signature of the emission enhancement seen at GeV energies is observed at VHEs. A careful statistical study based on the {\it{Fermi}} and H.E.S.S. lightcurves has been performed to check for this different behavior using the three-day overlap of the HE and VHE observations. Both the H.E.S.S. and {\it{Fermi}} data are divided into ``pre-flare" and ``flare" intervals. The HE flux ratio $\phi_{\rm flare}^{\rm HE}$/$\phi_{\rm pre-flare}^{\rm HE}$ is $\geq 9.2$ in the {\it{Fermi}} data, whereas a profile likelihood method applied to the VHE data yields a value $\phi_{\rm flare}^{\rm VHE}$/$\phi_{\rm pre-flare}^{\rm VHE}\,<3.5$ (99.7\% C.L.). The GeV and TeV emission may have therefore a different origin; if they were produced by the same processes, a flux enhancement of a similar magnitude in both energy bands would be expected. This conclusion is further supported by a joint fit to the HE and VHE spectra \cite{HESS2013}, which can not account for both emissions with any reasonable model, in particular when accounting for the reported {\it{Fermi}}-LAT upper limits at energies 1--100 GeV. One possible interpretation could be that VHE emission is produced by the shocked pulsar wind particles, the same responsible for the radio and X-ray emission, whereas the GeV flare could correspond to IC by the unshocked relativistic wind scattering off the companion's photon field as well as photons from a heated part of the Be circumstellar disk \cite{Khangulyan-2012}.

\subsection{HESS~J0632+057}
%-------------------------------------------------------------------------------------------------------------------

\noindent The point-like TeV source HESS~J0632+057 was discovered by H.E.S.S. during the Galactic Scan program \cite{Aharonian2007} and later confirmed as a TeV emitter through further VERITAS (\cite{Acciari-2009, Ong-2011}) and MAGIC (\cite{Aleksic2012}) observations. Its position is coincident with the massive B0pe-type star MWC 148 (HD 259440), for which optical spectroscopic observations have been recently obtained and confirm that HESS J0632+057 is indeed a binary system \cite{Casares2012}. HESS~J0632+057 is also coincident with the {\it{ROSAT}} source 1RXS~J063258.3+054857, the unidentified gamma-ray source 3EG J0634+0521, and has a point-like radio counterpart coincident with MWC~148 \cite{Skilton-2009}, which also shows variable extended radio emission \cite{Moldon-2011}. Its radio and X-ray properties \cite{Hinton2009, Falcone2010, Bongiorno2011, Rea-2011} resemble those found in LS~5039, LS~I~+61$^{\circ}$~303 and PSR~B1259-63. In contrast to these systems, however, no emission at MeV-GeV energies has been detected so far from the system, with a 99\% C.L. upper limit at energies $E>$100 MeV at the level of $F_{100} \leq 3.0\times 10^{-8}$ ph~cm$^{-2}$~s$^{-1}$, as reported by the \textit{Fermi}-LAT Collaboration \cite{Caliandro-2012}.

\noindent The total H.E.S.S. data-set on HESS~J0632+057 spans a wide time interval from 2004 to 2012, amounting to $\sim$~52 h of observations. Such a long observing time-range provides a good coverage of the $\sim 315$-days orbital periodicity of the system. The latest H.E.S.S. data correspond to the 2011/2012 campaign, with observations taken in December 2011 and February 2012, which provide a total of 8.2 h of data after standard quality cuts filtering. However, the expected maximum in the source light-curve could not be well-sampled in this occasion due to bad weather conditions, and most of the observations felt in the so-called X-ray ``dip"
phase right after the maximum, at orbital phases $\sim 0.4$ (\cite{Bongiorno2011, Bordas-Meier-2012}). The total data-set also includes H.E.S.S. observations corresponding to orbital phases poorly explored so far by any current Cherenkov telescope, in particular at phases in the range 0.6--0.9.  Data taken in March 2007, January 2008 and October 2009 amounting to $\sim 14$~h of observing time fall in this phase interval. Making use of the {\it{Model analysis}} technique \cite{Naurois2009}, the data provides a $> 7 \sigma$ significance detection at those phases. No trial factors are to be considered as the signal searched is based on the existence of a secondary peak in the source phase-folded X-ray lightcurve. Accounting nevertheless for a conservative trial factor estimation corresponding to a blind search, they would still provide a $> 6 \sigma$ significance detection.  A cross-check of these results, performed using a Hillas based analysis for the same run list, which also makes use of an independent calibration of the raw data, provides compatible results. A detailed report on these results, including an updated H.E.S.S data-set together with VERITAS and {\it{Swift}}-XRT long-term observations is in preparation.

%\begin{figure}
%\includegraphics[width=0.5\textwidth]{images/gamma2012_phsase_folded_HESS_alone.eps}
%\caption{\label{fig1}
%Figure.
%}
%\end{figure}

\subsection{HESS~J1018-589}
%-------------------------------------------------------------------------------------------------------------------

The {\it{Fermi}}-LAT collaboration has recently reported on the detection of the new gamma-ray binary candidate 1FGL\,J1018.6--5856 \cite{Ackermann-2012}. The periodic modulation of $P_{\rm orb} = 16.58 \pm 0.02$\,days observed in the {\it{Fermi}}-LAT data strongly suggests a binary nature of the source, with the O6V-type star 2MASS 10185560--5856459 proposed as the companion star \cite{Ackermann-2012}. Further indications of the its binary nature are provided by the observation of periodic X-ray and radio emission from a point-like source spatially coincident with 1FGL\,J1018.6--5856 \cite{Ackermann-2012}. At hard X-ray wavelengths, Li et al. (2011) \cite{Li-2012} found a possible INTEGRAL counterpart to the system, with also a hint of anticorrelation with respect to the {\it{Fermi}}-LAT phase-folded lightcurve. 

H.E.S.S. pointed towards the region around 1FGL\,J1018.6--5856 mainly during a multi-year observation campaign in 2007, 2008 and 2009 on Westerlund~2, leading to the detection of two distinct emission regions \cite{HESS2012a}. The first one consists of a diffuse emission region that extends towards the direction of PSR\,J1016--5857, the position of which is compatible with that of the pulsar, whilst the second one is a point-like source at the center of G284.3--1.8, which is spatially compatible with 1FGL\,J1018.6--5856. No flux variability could be derived in the 2007--2009 data. To further search for VHE variability/periodicity, new H.E.S.S. observations were performed in 2011-2012, improving the sampling of the system orbital phases. First evidences of flux variability have been found when these new data are included in the analysis. A detailed report on these results as well as a study on the VHE emission and absorption mechanisms at work in the source is in preparation.

%\begin{figure}
%\includegraphics[width=1\textwidth]{images/}
%\caption{\label{fig1}
%Figure.
%}

%
%
%\begin{figure}
%\includegraphics[width=1\textwidth]{images/PSRB1259_2004_2007_2011_LC_and_spectrum.eps}
%\caption{\label{fig3}
%Figure.
%}
%\end{figure}

\subsection{Eta~Carinae}
%-------------------------------------------------------------------------------------------------------------------

Eta~Carinae is a binary system located in the Carina Nebula, one of the most active HII region in our Galaxy. The system is composed by a Luminous Blue Variable (LBV) and an O- or B-type companion star, orbiting one another with a period of $\sim 5.5$~years in a highly eccentric orbit ($e \sim0.9$). The LBV displays a strong mass loss rate, $\dot{M} \geq 5 \times 10^{-4}\,M_{\odot}$~yr$^{-1}$, with a terminal wind velocity of $v_{\rm wind} \sim(500 - 700)$~km~s$^{-1}$. Together with the (less extreme) wind parameters of the companion star, the system outputs a total wind kinetic energy of about $10^{37}$~erg~s$^{-1}$ (see, e.g. \cite{Pittard-2002} and references therein). The collision region of the stellar winds may lead to efficient particle acceleration \cite{Reimer-2006}, which could be behind the high-energy emission observed from the system in X-rays \cite{Sekiguchi-2009, Leyder-2010} and HE gamma-rays \cite{Tavani2009b, Abdo2010}. Recently, a variable high-energy component has been found in the {\it{Fermi}}-LAT data, extending up to $\gtrsim 100$~GeV \cite{Farnier2011, Reitberger-2012}, further motivating the search for a possible VHE signal from the source.

H.E.S.S. observed the Carina region between 2004 and 2010 for a total of 33\,hours after standard quality selection cuts \cite{HESS2012b}. Observations were taken at zenith angles in the range 36$^{\circ}$ to 54$^{\circ}$, with a mean value of 39$^{\circ}$. The average pointing offset from the target position was 0.8$^{\circ}$. Data were analysed with the H.E.S.S. Standard Analysis software and made use of a Hillas-based Boosted Decision Trees method for an efficient suppression of the hadronic background component \cite{Ohm2009}. No significant signals of VHE $\gamma$-ray emission from Eta~Carinae or the Carina Nebula have been obtained from the H.E.S.S. observations. A 99\% confidence level upper limit on the integral $\gamma$-ray flux above 470\,GeV at the level of $7.7\times10^{-13}$~ph~cm$^{-2}$~s~$^{-1}$ and $4.2\times10^{-12}$~ph~cm$^{-2}$~s~$^{-1}$  have been derived for the central system and the surrounding nebula, respectively. The non-detection, together with the spectral properties of the high-energy component observed in
the {\it{Fermi}}-LAT data, implies the presence of a cut-off at energies $\lesssim 1$~TeV, caused either by a cut-off in the accelerated particle spectrum or as a consequence of a strong $\gamma$-$\gamma$ absorption in the inner regions of the binary system in either leptonic or hadronic scenarios \cite{Farnier2011, Bednarek-2011b}.

\section{Concluding remarks}

H.E.S.S. observations are providing new and exciting insights on most of the currently known gamma-ray binaries. These results
unambiguously disentangle the HE and VHE emission in the pulsar-wind/disk interaction region in PSR~B1259-63/LS 2883 during a recently detected HE flare. They reveal a periodic detection at VHE of HESS~J0632+057 at orbital phases close to the maximum of the X-ray lightcurve, whereas also a significant VHE emission at later orbital phases is observed. H.E.S.S. observations suggest a VHE counterpart in the extended VHE source HESS~J1018-589 to the recently discovered gamma-ray binary 1FGL~J1018.6-5856. Finally, upper limits obtained at VHEs on Eta~Carinae, together with the {\it{Fermi}}-LAT detection of a variable high-energy component in the source spectrum, imply the presence of a cut-off in the particle energy distribution and/or the signatures of strong attenuation due to $\gamma$-$\gamma$ absorption close to the wind-wind collision region.

\vspace{0.5cm}

\noindent Further H.E.S.S. observations of binary systems are foreseen in the near future. They will make use of the recent H.E.S.S. II
upgrade, providing an improved sensitivity and, for the first time, the possibility to use the Cherenkov technique
to detect sources down to $\sim 30$~GeV energy thresholds in comparatively short ($\sim$ few $\times 10$~h) exposure times.

%
%{\bf IMPORTANT:} Do not include page numbers on your paper. Page
%numbers will be assigned by the publisher.
%
%{\bf REMINDER:} Please, remember to choose the {\bf short title} for your contribution (line 14 in the .tex file).
%
%\section{Latex instructions}
%
%
%
%
%
%
%\subsection{Figures}
%
%Single column figures and images as in figure \ref{simp_fig}
%preferred. However, large figures can be made to span two columns. If you have to introduce two column
%figures(e.g., figure \ref{wide_fig}), please verify that they do not
%overflow the margins of the text.
%
%
% \begin{figure}[t]
%  \centering
%  \includegraphics[width=0.4\textwidth]{icrc2013-template-02}
%  \caption{Simple figure example. Rio Scenarium Restaurant / Bar / Nightclub. Venue of the Conference Dinner.}
%  \label{simp_fig}
% \end{figure}
%
%\subsection{Tables}
%
%Tables can be done at one (Table \ref{table_single}) and two columns. \\
%
%\begin{table}[h]
%\begin{center}
%\begin{tabular}{|l|c|c|}
%\hline Month & Temperature ($^\circ$C) & Precipitation (mm) \\ \hline
%June   & 22  & 81.3 \\ \hline
%July   & 22  & 55.9 \\ \hline
%August & 22  & 50.8 \\ \hline
%\end{tabular}
%\caption{Weather conditions in the city of Rio de Janeiro. Average temperature and precipitation in each month.}
%\label{table_single}
%\end{center}
%\end{table}

\vspace*{0.5cm}
\footnotesize{{\bf Acknowledgment:}{The support of the Namibian authorities and of the University of Namibia in facilitating the construction and operation of H.E.S.S. is gratefully acknowledged, as is the support by the German Ministry for Education and Research (BMBF), the Max Planck Society, the French Ministry for Research, the CNRS-IN2P3 and the Astroparticle Interdisciplinary Programme of the CNRS, the UK Science and Technology Facilities Council (STFC), the IPNP of the Charles University, the Polish Ministry of Science and Higher Education, the South African Department of Science and Technology and National Research Foundation, and by the  University of Namibia.We appreciate the excellent work of the technical support staff in Berlin, Durham, Hamburg, Heidelberg, Palaiseau, Paris, Saclay, and in Namibia in the construction and operation of the equipment. This research has made use of the NASA/IPAC Infrared Science Archive, which is operated by the Jet Propulsion Laboratory, California Institute of Technology, under contract with the National Aeronautics and Space Administration. PB has been supported by grant DLR 50 OG 0601 during this work. PB also acknowledges the excellent work conditions at the \textit{INTEGRAL} Science Data Center.}}

\end{document}